\documentclass[showpacs,amsmath,twocolumn,showkeys,pra]{revtex4-1}

\usepackage{dcolumn}    
\usepackage{bm}         
\usepackage{ifpdf}
\usepackage{amssymb,lineno,amsfonts}
\usepackage{graphicx}   
\usepackage{bbm}
\usepackage{mathrsfs}
\usepackage{upgreek}
\usepackage{mathtools}
\usepackage{epstopdf}
\usepackage{setspace}
\usepackage{hyperref}
\usepackage[matrix,frame,arrow]{xypic}
\usepackage{rotating}
\usepackage{float}
\usepackage{natbib}
\usepackage{color} 
\definecolor{med-blue}{RGB}{25,25,112} 
\hypersetup{colorlinks, linkcolor={red},citecolor={blue}, urlcolor={blue}}

\usepackage{color} 
\definecolor{med-blue}{RGB}{25,25,112} 
\hypersetup{colorlinks, linkcolor={red},citecolor={blue}, urlcolor={blue}}
\newcommand{\ket}[1]{|{#1}\rangle}

\newcommand{\matrixel}[3]{\langle {#1} | {#2} | {#3} \rangle}
\newcommand{\outpr}[1]{|{#1} \rangle  \langle {#1}|}

\begin{document}
\title{Quantum simulations of a particle in one-dimensional potentials using NMR}
\author{Ravi Shankar, Swathi S. Hegde, and T. S. Mahesh}
\email{mahesh.ts@iiserpune.ac.in}
\affiliation{Department of Physics and NMR Research Center, Indian Institute of Science Education and Research, Pune 411 008, India}

\begin{abstract}
{
A classical computer simulating Schrodinger dynamics of a quantum system requires
resources which scale exponentially with the size of the system, and is 
regarded as inefficient for such purposes. However, a quantum computer made up of
a controllable set of quantum particles has the potential to efficiently simulate
other quantum systems of matching dimensions. 
In this work we studied quantum simulations of single particle Schrodinger equation 
for certain one-dimensional potentials. In particular, we report the following
cases: 
(i) spreading of wave-function of a free-particle, 
(ii) evolution of a particle in a potential-well, and
(iii) reflection of a particle from a potential-barrier.
  Using a five-qubit NMR system, we achieve
space discretization with four qubits, and the other qubit is used for 
preparation of initial states as well as measurement of spatial probabilities.
The experimental relative probabilities compare favourably with the theoretical
values, thus effectively mimicking a small-scale quantum simulator.
}
\end{abstract}

\maketitle

\section{Introduction}
We have reached a stage in digital computing, conforming to Moore's law, where quantum effects are more pronounced and serve as a hindrance to transistor performance \cite{moore}. The way forward, as Feynman suggested, is to use these effects to our advantage in building a computer - a quantum computer \cite{feynman}. 
The fundamental working of any quantum computer, if built, boils down to initializing a quantum state, transforming it by a desired unitary operator, and finally performing an efficient readout of the final state. It has been established theoretically that quantum computers have the capability to
alter the complexity of certain computational problems from exponentials to polynomials.  A well-known example
is Shor's algorithm for prime-factorization \cite{shor}. Another important example is simulating the dynamics of 
quantum systems \cite{feynman}.
Dynamics of a quantum particle in a potential is governed by Schrodinger equation. 
Simulating the dynamics is carried out by assigning a state vector to the system, and transforming
it using unitary operators.  For a system consisting of  $n$ mutually interacting spin-1/2 particles
the dimension of the state vector increases as $2^n$ and that of unitary operator as $2^n \times 2^n$.
Thus the number of variables in the problem increases exponentially with the size of the system.
Nevertheless, as Feynman noted, a quantum system can be efficiently simulated with the help of another controllable quantum system - a quantum simulator \cite{feynman,buluta}. 

Extensive studies have been made on quantum simulations using both 
theory and experiments.  After Feynman popularized the concept of quantum simulators,
the idea was further studied by Lloyd and Braunstein \cite{Lloyd,braunstein}. 
Later quantum simulations of various problems were investigated in detail.
For example, Farhi et al have studied simulations of quantum walk \cite{farhi}.
Cory and co-workers have simulated the dynamics of truncated quantum
harmonic and anharmonic oscillators using nuclear magnetic resonance (NMR) \cite{cory}.
Laflamme and co-workers have simulated quantum many-body problems using similar 
techniques \cite{laflamme}.
Wineland and co-workers have realized an ion-trap quantum simulator
and demonstrated its application as a non-linear interferometer \cite{wineland}.
Jingfu et al have simulated a perfect state transfer in spin chains using 
Heisenberg XY interactions on an NMR system \cite{jingfu}.  
White and co-workers have simulated molecular hydrogen and experimentally
obtained its energy spectrum using photonic quantum computer technology \cite{white}.
Du and co-workers have also obtained the ground state energy of hydrogen
molecule by an NMR quantum simulator \cite{du}.
Oliveira and co-workers have simulated magnetic phase transitions by writing
electronic states of a ferromagnetic system onto nuclear spin states \cite{oliveira}.
More recently, Koteswara et al have used an NMR quantum simulator to establish
multipartite quantum correlations as an indicator of frustration in Ising spin system
\cite{kota}.
The applications of
spin-qubits as quantum simulators are discussed in a review by Peng and Suter \cite{suter}.

In this article we report the experimental demonstration of time-evolution of a quantum particle under various potentials using NMR techniques on a 5-qubit system. As described in the next section, the quantum simulation 
of a particle in a potential can be decomposed into two parts: the evolution under potential energy
in position basis, and the evolution under kinetic energy in momentum basis.  The change of basis
is achieved by Fourier transform and finally the probabilities are measured in the position basis.
In our work, we utilize four qubits to discretize the position space.  We also utilize an ancilla qubit 
to (i) prepare an initial state and (ii) directly encode the spatial probabilities onto 
the ancilla spectral lines.  
Therefore the need of a quantum state tomography 
and associated post-processing of data, is avoided.

In the following section we outline the theory of quantum simulations of a particle in a 
one-dimensional potential. In section III we describe the experimental methods in simulating
a few potentials and discuss the results.  Finally we
conclude in section IV.

\section{Theory}
The dynamics of a quantum particle of mass $\mu$ in a one-dimensional potential $V(x)$ is 
governed by the time-dependent Schrodinger equation,
\begin{equation} \label{xse}
i\hbar\frac{\partial \ket{\psi}}{\partial t} = {\cal H}\ket{\psi},
\end{equation}
where 
$\ket{\psi}$ is the state vector and
\begin{equation} \label{hprob}
{\cal H} = {\cal H}_0 + V(x)
\end{equation} 
is the Hamiltonian in position basis ($x$-basis).  
Here 
${\cal H}_0 = \frac{-\hbar^2}{2\mu} \frac{\partial^2 }{\partial x^2}$
represents the kinetic part of the Hamiltonian.
The mass $\mu$ and the reduced Planck's constant $\hbar$
are set to unity from here onwards.  Further we use the position basis
to encode the initial state and to measure the probabilities of the 
final state.  If the system is represented by an initial wave-function
$\psi(x,t)$ at a time instant $t$, after an evolution over a small time 
duration $\delta t$, the wave-function is given by
\begin{equation} \label{sesoln}
\psi(x,t+\delta t) = e^{-i {\cal H} \delta t} \psi(x,t).
\end{equation}

We choose a domain of length $L$ to which the spread of the 
wave-function is confined in $x$-basis. We utilize an $n$-qubit register
to encode discrete positions 
\begin{equation}
x_j = x_0+j\Delta x.
\label{xj}
\end{equation}
Here 
the index $j$ runs from 0 to $N-1$, with $N=2^n$,
$x_0=-L/2$ and $x_{N-1} = +L/2$ forming the terminals, and
the resolution of discretization, $\Delta x = L/(N-1)$, is
limited by the size of the register.
The state vector can now be expressed in position basis as
\begin{equation}
\ket{\psi(t)} = \frac{1}{\sqrt{N}}\sum_{j=0}^{N-1} \psi(x_j,t) \ket{j}
\end{equation}
where $\{\ket{j}\}$  represents the computational basis formed by the $n$-qubit register \cite{benenti}.

Since the two parts, ${\cal H}_0$ and $V$, of the Hamiltonian do not commute, one can 
utilize Trotter's formula
\begin{equation}
e^{-i {\cal H} \delta t} = e^{-i V \delta t/2} e^{- i {\cal H}_{0} \delta t} e^{-i V \delta t/2} + O(\delta t^3)
\end{equation}
to decompose the total evolution over a duration $\delta t$ 
into independent evolutions under potential and kinetic
parts.

In the discretized position basis, the evolution under the potential part $V(x)$ leads to
the transformed wave-function
$e^{-i V(x_j) \delta t} \psi(x_j,t)$.
This is equivalent to applying a diagonal unitary operator to the state-vector.
Such a transformation can be implemented efficiently on a quantum computer
\cite{chuangbook,benenti}.  However, the second part, i.e., the kinetic energy part
in the Hamiltonian (Eq. [\ref{hprob}]) is non-diagonal in the position basis, and
has no efficient decomposition.  The trick involves applying discrete Fourier transform ${\cal F}$
to transform the wave-function into momentum basis \cite{ftref}, i.e.,
\begin{equation} \label{eqft}
\widetilde{\psi}(k_l, t) =  \frac{1}{\sqrt{2\pi}} \sum_{j=0}^{N-1} \psi(x_j, t) e^{ik_l x_j} \Delta x.
\end{equation}
In the discretized momentum basis
($k$-basis), 
\begin{equation}
k_l = k_0+l \Delta k 
\label{kl}
\end{equation}
is the $l$th point, with
$\Delta k = 2\pi/L$, and $k_0 = -(N-1)\Delta k /2$ and $k_{N-1} = +(N-1)\Delta k /2$ forming the terminals.
In this basis, the kinetic-energy part $\widetilde{\cal H}_{0}$ becomes diagonal, i.e.,
$\matrixel{k_j}{\widetilde{\cal H}_{0}}{k_l} = \delta_{jl}k_l^2/2$,
where $\delta_{jl}$ is the Kronecker delta function.
Therefore the transformed wave-function is of the form
$e^{-i (k_l^2/2) \delta t} \widetilde{\psi}(k_l,t)$.
Again the methods of efficient decompositions can be used to apply
the diagonal operator
\cite{wiesner, zalka, strini, benenti}.  

Thus the net evolution over a small duration $\delta t$ can be approximated as 
\begin{eqnarray}
e^{-i {\cal H} \delta t} \approx e^{-i V \delta t/2} {\cal F}^{-1} e^{- i \widetilde{\cal H}_{0} \delta t} {\cal F} e^{-i V \delta t/2}.
\end{eqnarray}
Hence a single time-step ($\delta t$) evolution of $\ket{\psi}$ is achieved by the application of
operators $U_V = e^{-i V \delta t/2}$, ${\cal F}$, $\widetilde{U}_{0} = e^{- i \widetilde{\cal H}_{0} \delta t}$, and ${\cal F}^{-1}$ in the sequence given above.

In the next section we describe the experimental methods employed for 
the quantum simulations of certain one-dimensional potentials
and discuss the results obtained.

\section{Experiments}
\begin{figure}
\centering
\includegraphics[trim=3cm 3.5cm 0.5cm 3cm, clip=true,width=9cm]{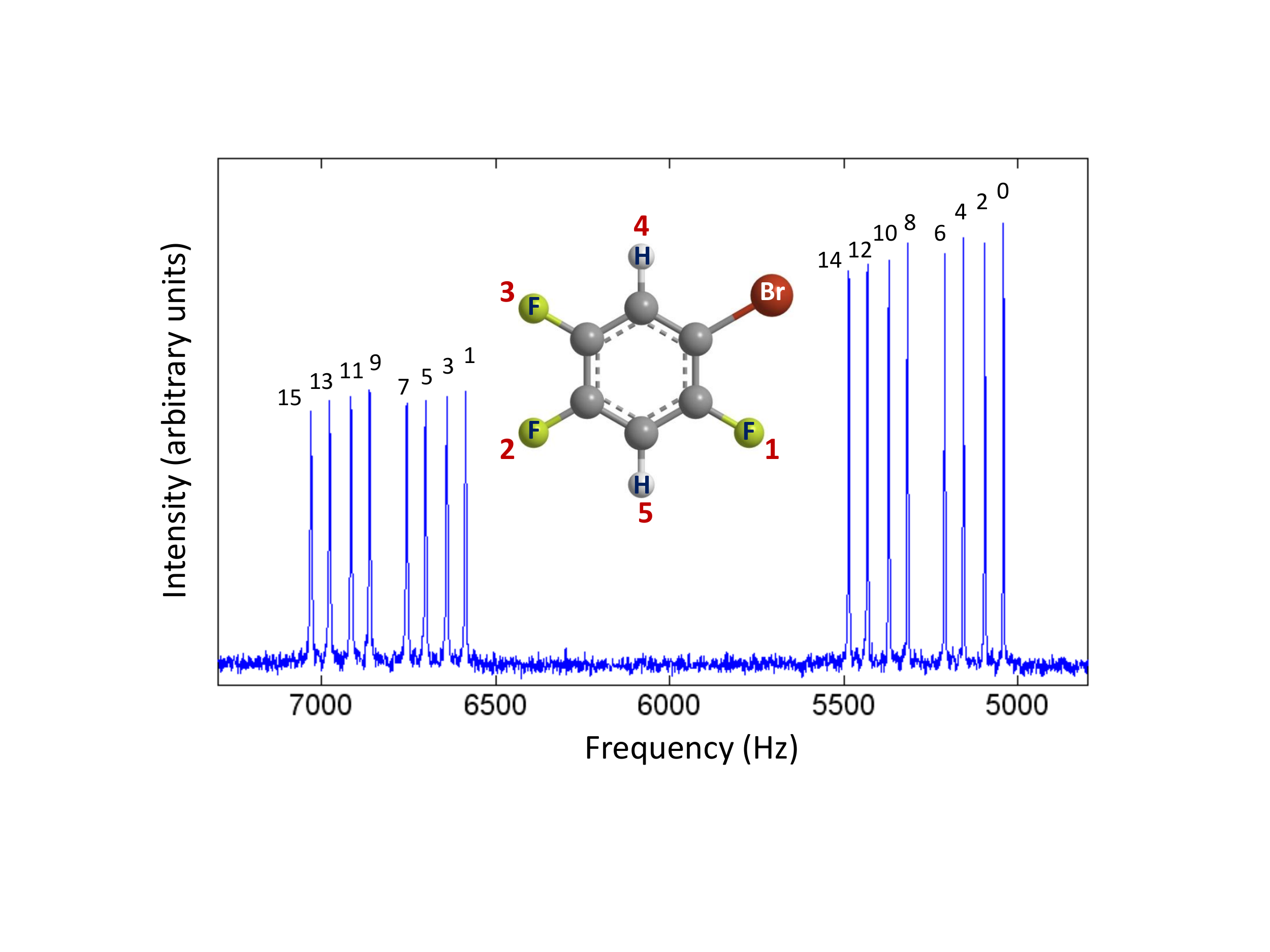}
\caption{Molecular structure of 1-bromo-2,4,5-trifluorobenzene and 
$^{19}$F spectrum of F$_1$ spin.  The state labels ($\ket{j}$) 
shown at the top of each transition are 
obtained through Z-COSY experiments and are consistent with the spin-states of the register
(qubits 2 to 5).
} \label{ref}
\end{figure}
In our experiments we utilize three $^{19}$F spins and two $^1$H spins
of 1-bromo-2,4,5-trifluorobenzene partially oriented
in liquid crystal N-(4-methoxybenzaldehyde)-4-butylaniline 
(MBBA). 
The molecular structure and the labelling of qubits are shown in
the inset of Fig. \ref{ref}.
All the experiments are carried out on a 500 MHz Bruker
UltraShield spectrometer at an ambient temperature of 300 K.
The effective couplings  in such a system is due to
scalar interactions (J-couplings) as well as 
partially averaged dipolar interactions (DD-couplings).  
The values of resonance frequencies
in a doubly rotating frame, 
longitudinal relaxation time constants (T$_1$),
effective transverse relaxation time constants (T$_2$),
and strengths of effective couplings ($D_{ij}$)
are tabulated below:
\begin{center}
\begin{tabular}{|c|c|c|c|}
\hline
Spin  & $\nu_i$ & $T_1$ & $T_2^*$\\
  number    &   (Hz)     & (s)   & (ms) \\
\hline
1 & 6029 & 0.7 & 65-125\\
2 & -3680 & 0.4 & 45-65\\
3 & -6743 & 0.5 & 45-65\\
4 & 50 & 1.4 & 150\\
5 & 29 & 1.3 & 150\\
\hline
\end{tabular}
\quad
\begin{tabular}{|l|l|}
\hline
\multicolumn{2}{|c|}{$D_{ij}$ (Hz)} \\
\hline
$D_{12}$ = 277 & $D_{24}$ = 106\\
$D_{13}$ = 116 & $D_{25}$ = 1270\\
$D_{14}$ = 54 & $D_{34}$ = 1532\\
$D_{15}$ = 1556 & $D_{35}$ = 55\\
$D_{23}$ = -26 & $D_{45}$ = -7.6\\
\hline
\end{tabular}
\end{center}
As evident from the above table, all the 80 transitions in the oriented 5-spin
system are fully resolved.  We utilize the first qubit as ancilla and rest 
(i.e., qubits 2 to 5) as the four-qubit quantum register, whose computational 
states encode the 16 discrete points in position as well as momentum 
bases.  Further, the ancilla has 16 well-resolved transitions, each of which 
corresponds to one particular qubit-state of the register (see Fig. \ref{ref}).

Since the chemical shift difference $\vert \nu_j-\nu_i \vert$ 
in each pair of spins $(i,j)$
is much higher than the intra-pair effective coupling strength ($D_{ij}$), the 
Hamiltonian can be approximated to the form \cite{cavanagh}
\begin{eqnarray}
&&{\cal H}_{\mathrm{int}} = -\pi \sum\limits_{i=1}^{n}\nu_i \sigma_i^z 
+ \frac{\pi}{2} \sum\limits_{\substack{i,j=1 \\ i<j}}^{n} D_{ij} \sigma_i^z\sigma_j^z.
\label{hint}
\end{eqnarray}

\begin{figure}
\includegraphics[width=7cm]{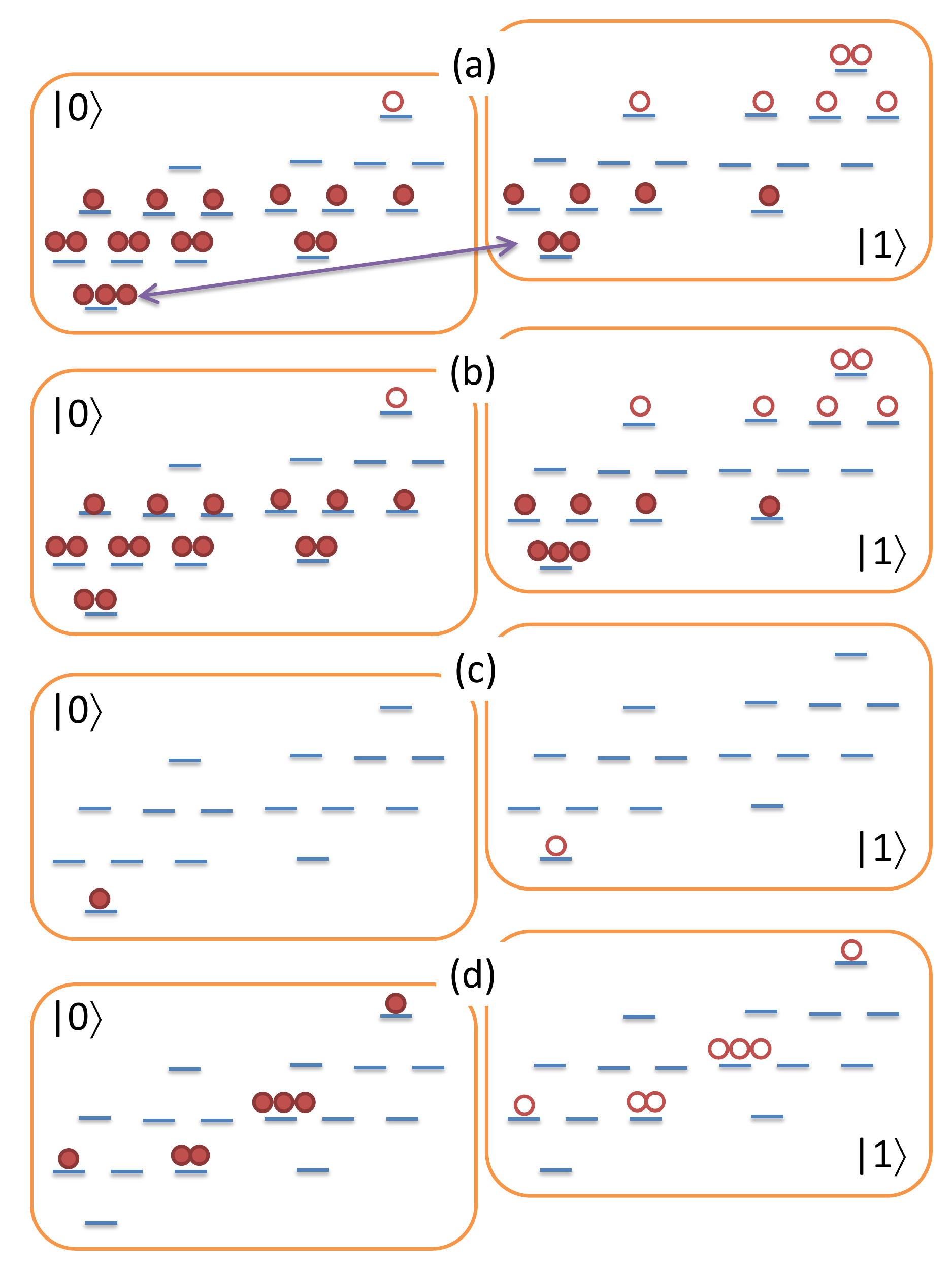}
\caption{The energy-levels and representative deviation population distributions
in the two subsystems corresponding to $\ket{0}$ and $\ket{1}$ states of the ancilla qubit.
The filled circles indicate positive deviations (above the background population) and 
the empty circles indicate negative deviations (below the background population).  
Here following situations are illustrated
(a) equilibrium distribution, (b) after inverting a transition (as shown in (a)), (c)
difference between (a) and (b) forming a POPS, and 
(d) an example of intra-subsystem redistribution of 
the differential populations after applying an unitary (with identity on ancilla).
\label{pops} } 
\end{figure}

The first step in quantum simulation typically involves preparation of an initial pure state
which is difficult using conventional NMR techniques. Instead a pseudopure state (PPS), with the density matrix
\begin{eqnarray}
\rho(0) & \ = \ & \frac{1}{2^{n}}(1-\epsilon) \mathbbm{1} + \epsilon |\psi\rangle\langle\psi|,
\end{eqnarray}
isomorphous to a pure state $\outpr{\psi}$, can be prepared. 
There are a number of methods available for this purpose 
\cite{coryspavg,knilltavg,anilll,knillcat,soumyasinglet}.
Here we utilize the pair of pseudopure states (POPS) method
described by Fung \cite{fung}.
In this method, as illustrated in Fig. \ref{pops},  the full-system basis is divided into two sub-systems 
based on the ancilla states.  
The preparation consists of a pair of experiments, one starting
from equilibrium population distribution $\{p_i\}$ (Fig. \ref{pops}(a)), 
and the other starting from an initial distribution $\{q_i\}$ obtained by 
inverting a single transition of ancilla (Fig. \ref{pops}(b)).  
Assuming linear response, the difference between the signals obtained
in the two experiments corresponds to an initial state which is the difference 
of the initial population distributions, i.e., $\{p_i-q_i\}$ (Fig. \ref{pops}(c)).
For example, if the ancilla transition between the levels $\ket{00000}$ and $\ket{10000}$
is inverted, one obtains $\ket{0000}$ POPS (see Fig.  \ref{pops}(c)), 
which encodes the quantum system localized at position $\ket{j=0}$.

If the subsystems are not allowed to mix during the computation,
(i.e., if no pulses are applied on the ancilla qubit) they undergo independent, 
simultaneous, and identical evolutions.  Therefore the initial populations
of the subsystems are redistributed within each subsystem in an identical 
manner (Fig. \ref{pops}(d)), such that the differential deviation populations 
are same, but are of opposite signs in the two subsystems.
In this work, the spatial probabilities are encoded by these  differential populations $\{p_i-q_i\}$ 
and are reflected in the intensities of the ancilla spectral lines.

\begin{figure}
\centering
\includegraphics[trim=6cm 3.5cm 0.2cm 2cm, clip=true,width=10cm]{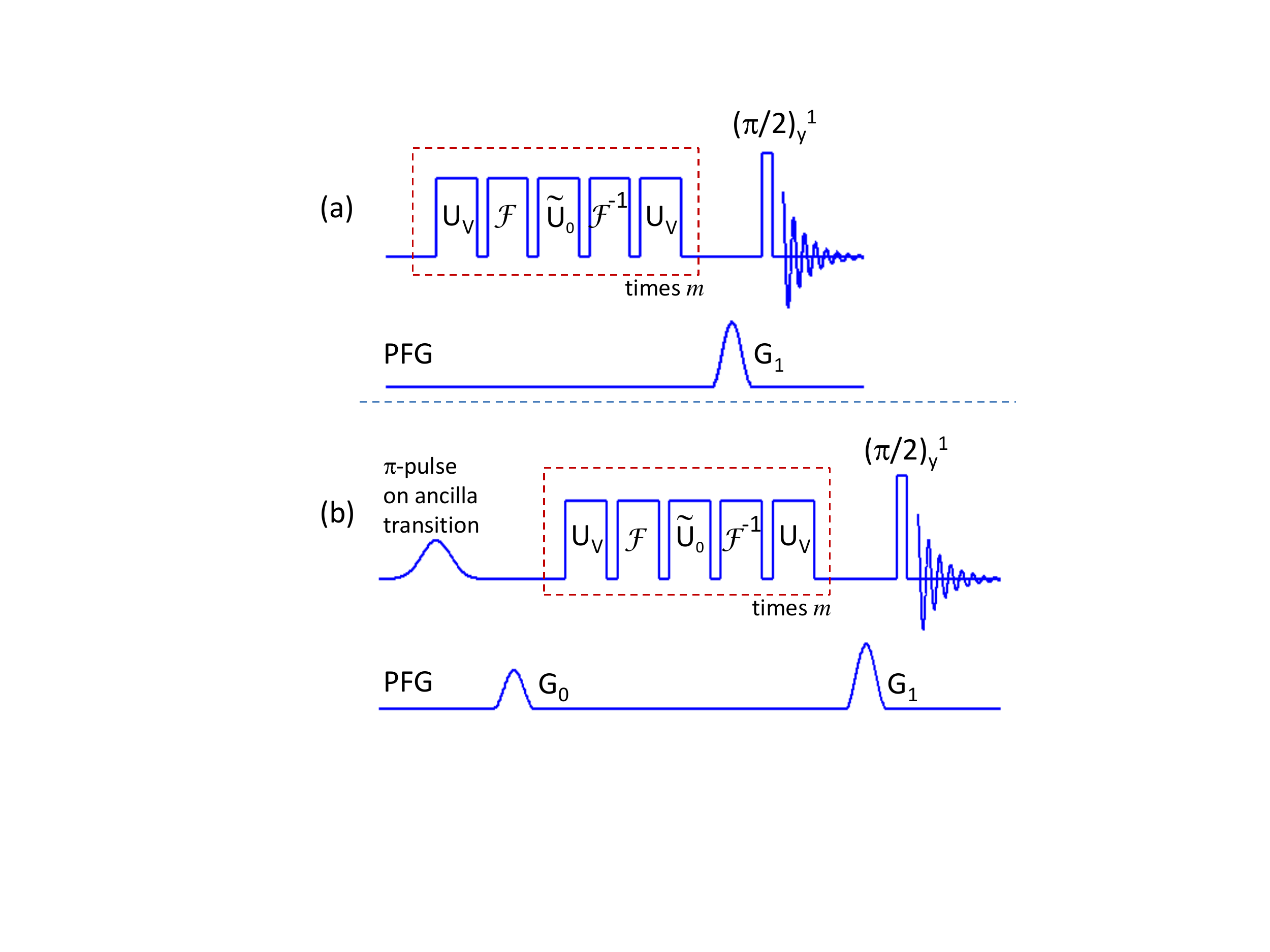}
\caption{
NMR pulse sequences used for simulating the quantum dynamics.  
The sequence (b) differs from (a) only by the initial transition selective
inversion followed by a PFG (G$_0$).  
The PFG G$_1$ is used to destroy all the transverse magnetization before
the final $(\pi/2)_y^1$ detection pulse on the ancilla.  The operators in
the dashed box, corresponding to the Schrodinger evolution, are repeated $m$ times to 
achieve an evolution over duration $m\delta t$.
} 
\label{ppg}
\end{figure}

The pulse sequence for the simulation experiments are illustrated in Fig. \ref{ppg}.
To measure the spatial probabilities after a time interval $\delta t$, 
we carry out two experiments 
(i) apply $e^{-i{\cal H}\delta t}$ to equilibrium state $\rho(0)$, destroy the
transverse magnetization by using a pulsed field gradient (PFG), 
and read the signal by applying a linear detection
pulse ($(\pi/2)_y^1$ pulse on ancilla) (see Fig. \ref{ppg} (a)), and
(ii) invert the desired ancilla transition, apply $e^{-i{\cal H}\delta t}$,
 destroy the transverse magnetization, and again read the signal by applying the 
linear detection pulse (see Fig. \ref{ppg} (b)).
The difference between the signals obtained in these two experiments directly
encode the spatial probabilities. Thus the ancilla qubit provides a
direct spectral information on the dynamics of the quantum system.
In order to study the dynamics at a set of regularly spaced time instants, we repeat the
evolution pulses shown in dashed boxes in Fig. \ref{ppg}.

The initial transition selective $\pi$ pulse in Fig. \ref{ppg}(b) is realized 
by a long Gaussian RF pulse of duration 30 ms.
Other RF pulses used in the above pulse sequences are constructed using
GRadient Ascent Pulse Engineering (GRAPE) technique and are designed 
to be robust against RF inhomogeneity.
The operators shown in the dashed box, for each potential, in Fig. \ref{ppg} are constructed 
as a single GRAPE pulse of duration 24 ms.  
The fidelities (averaged over inhomogeneous spatial distribution of
RF amplitudes) of these pulses were better than 0.95.  
The final selective $(\pi/2)_y^1$ pulse on the ancilla was of duration 
500 $\upmu$s and of average fidelity better than 0.99.

\begin{figure}[b]
\centering
\includegraphics[trim=0cm 2cm 0cm 0cm, clip=true,width=8cm]{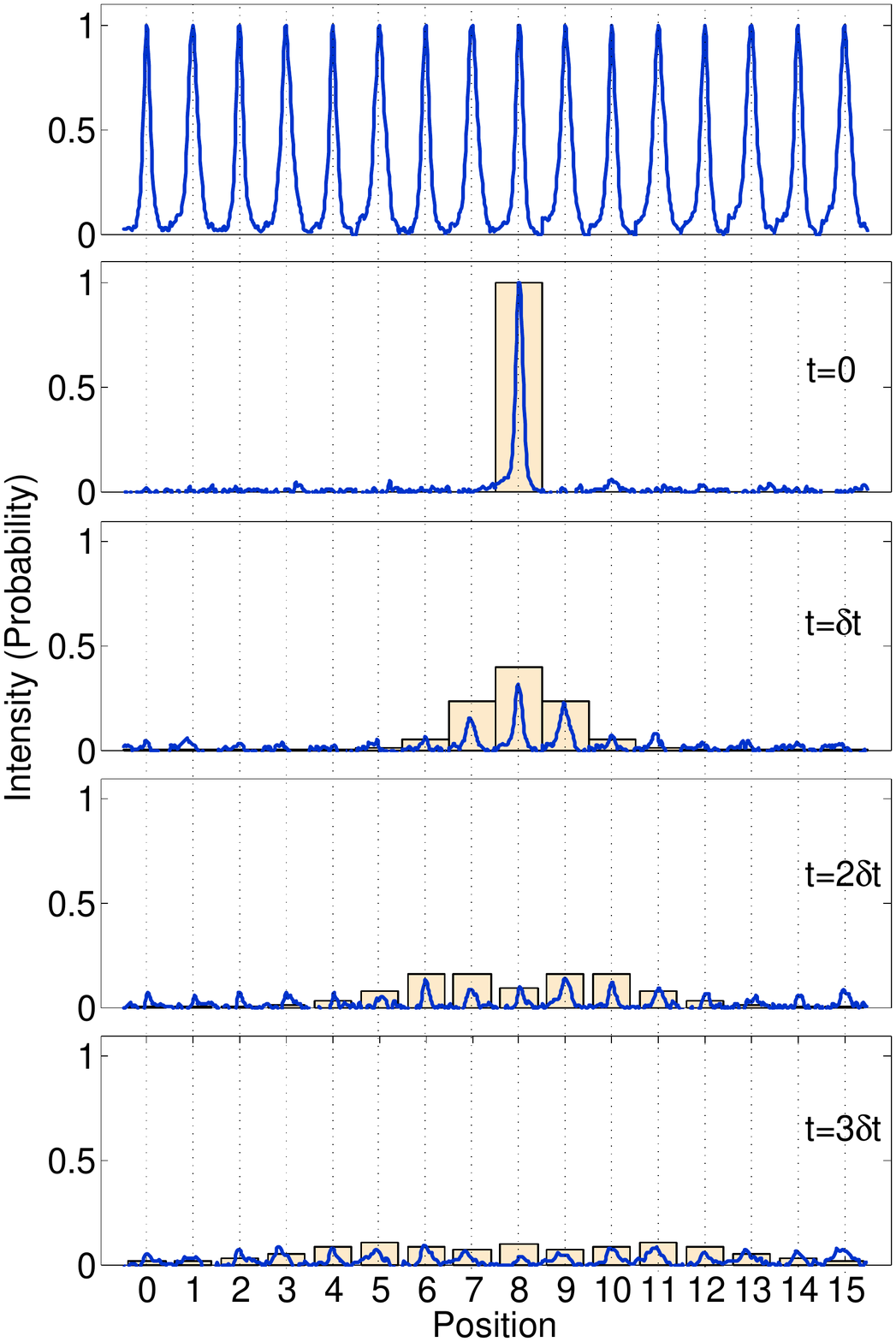}
\caption{
Top trace is the reference spectrum of ancilla obtained by rearranging normalized 
peaks from Fig. \ref{ref} in the increasing order of spectral labelling.  Here
position is in units of $\Delta x$.
The spectra at $\delta t = 0$ corresponds 
to the $\ket{j=8}$ POPS.  Other spectra correspond to different 
evolution intervals as indicated.  Shaded bars shown behind the spectral
lines indicate expected probabilities.
} \label{v0}
\end{figure}

\begin{figure}[t]
\centering
\includegraphics[trim=0cm 2cm 0cm 0cm, clip=true,width=8cm]{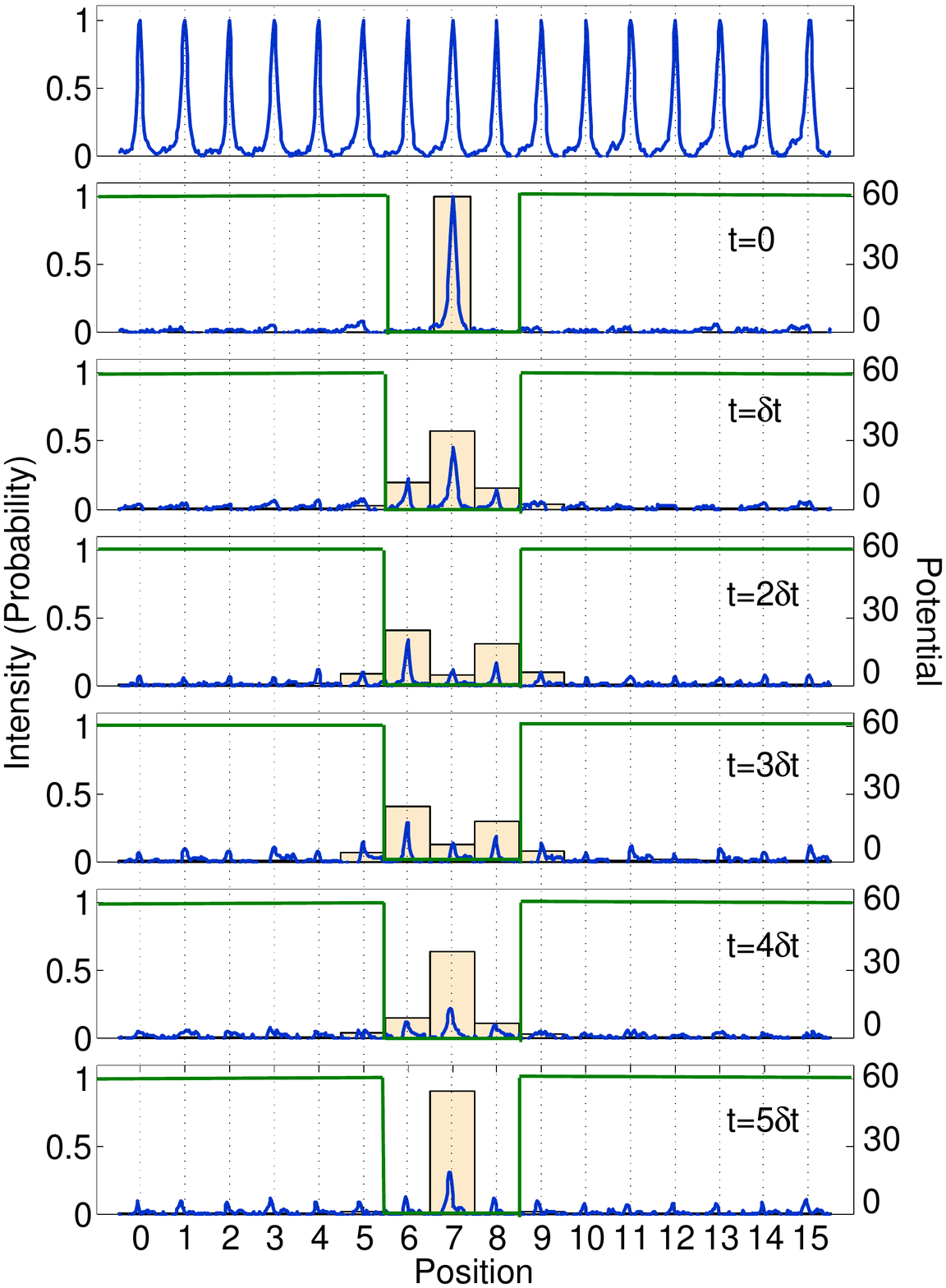}
\caption{
Reference spectrum (top trace) and spectra corresponding to the 
simulation of particle in a well for sequential intervals of time
as indicated.
The spectra at $\delta t = 0$ corresponds 
to the $\ket{j=7}$ POPS.   Shaded bars shown behind the spectral
lines indicate expected probabilities. Potential is indicated by
a solid line.
} \label{well}
\end{figure}

\begin{figure}
\centering
\includegraphics[trim=0cm 2cm 0cm 0cm, clip=true,width=8cm]{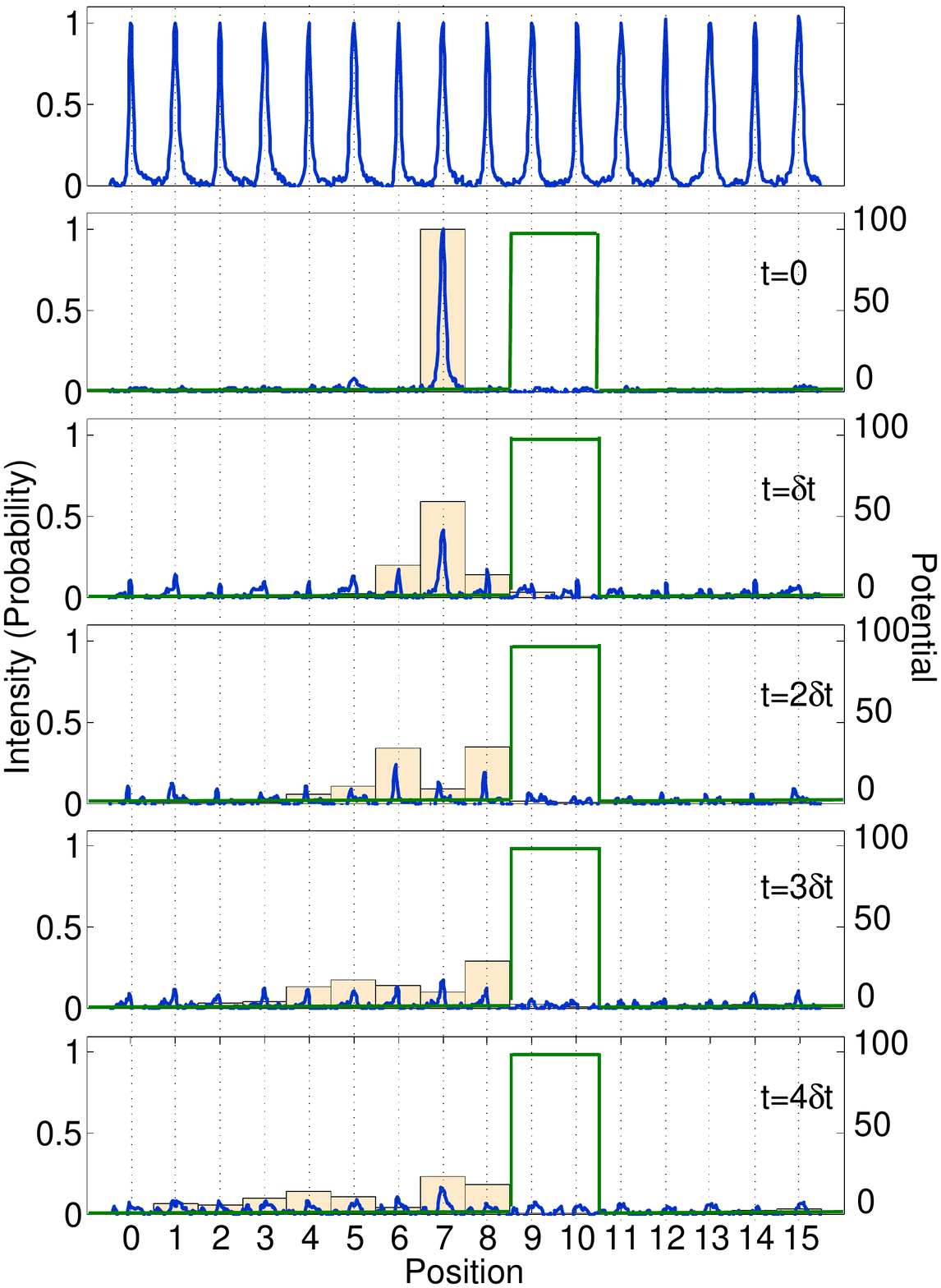}
\caption{
Reference spectrum (top trace) and spectra corresponding to the 
simulation of reflection from a barrier for sequential intervals of time
as indicated.
The spectra at $\delta t = 0$ corresponds 
to the $\ket{j=7}$ POPS.   Shaded bars shown behind the spectral
lines indicate expected probabilities. Potential is indicated by
a solid line.} \label{bar}
\end{figure}

We simulated the quantum dynamics of a particle starting from rest 
(with zero momentum expectation value) in the following three cases.
\begin{itemize}
\item[(i)]
{Free-particle (Fig. \ref{v0}): $V(x) = 0$ for all values of $x$ in 
a lattice of length $L = 8$. 
The particle starts from rest at $x=4/15$ ($j=8$ in Eq. (\ref{xj})) and the evolution interval is $\delta t = \pi/20$.
}
\item[(ii)]{
Particle in a well (Fig. \ref{well}): $V(x) = 0$ for $-2/5 \le x \le 2/15$ (i.e., $6  \le j \le 8$), and 
$V(x) = 60$ elsewhere in a lattice of length $L=4$. 
The particle starts from rest at $x=-2/15$ ($j=7$) and the evolution interval is $\delta t = {\pi}/{100}$.
}
\item[(iii)]{Reflection from a barrier (Fig. \ref{bar}): $V(x) = 100$ at $2/5 \le x \le 2/3$ (i.e., $9 \le j \le 10$) 
and $V(x) = 0$ elsewhere in a lattice of length $L=4$. 
The particle starts from rest at $x=-2/15$ ($j=7$) and the evolution interval is $\delta t = {\pi}/{100}$.
}
\end{itemize}
It can be noticed that units of $V$ and $L$ are set by the conditions $\hbar = 1$ and mass $\mu=1$.
The parameter values in the above are chosen considering the contrast of spatial probabilities
at different instants of time.

In the following we discuss each of these cases.  
For the case of free-particle, there is no need to apply $U_V$ mentioned in
the pulse-sequence shown in Fig. \ref{ppg}. The GRAPE pulse implementing
the propagator ${\cal F}^{-1}\widetilde{U}_0{\cal F}$ had an average
fidelity of 0.98.  The experimental results are shown in Fig. \ref{v0}.
The ancilla spectral lines normalized into equal intensities and  
ordered in the increasing values of transition-labels $(j)$ are shown in 
the top trace of Fig.\ref{v0}. These spectral lines are used as references
for normalizing spectral lines in the other traces of the figure. 
The spectrum corresponding to $t=0$ is obtained from $\ket{j=8}$ POPS.  
The remaining traces show the spectra obtained after subsequent intervals
of evolution (up to $3 \delta t$).  The expected spatial probabilities under an
ideal reproduction of GRAPE pulses are 
shown as shaded bars behind the spectral lines.  The spreading of
the wave-function in zero-potential is clearly manifested in both theoretical
and experimentally simulated plots.  Ideally peak-heights of the spectral
lines and the bars should match.  Experimentally however, the 
spectral lines are generally of lower intensities as a result of decoherence.
The overall duration of the pulse-sequence for simulating $3\delta t$ evolution
is about 72 ms which is comparable to the $T_2$ values of the various spins in
the molecule.  Another major challenge is the gradual fluctuations in
the residual dipolar couplings due to the local changes in the 
order parameter of the liquid crystal.  This resulted not only in the 
line-broadening of spectral lines, but also temporal modulations in the 
Hamiltonian parameters.
Other sources of errors like RF inhomogeneity, static-field inhomogeneity, 
and non-linear reproduction of the GRAPE
pulses by the RF coils also contribute to the deviations of the experimental
spectra from theory.  

\begin{figure}[b]
\centering
\includegraphics[trim=0cm 3.2cm 0cm 3cm, clip=true,width=9cm]{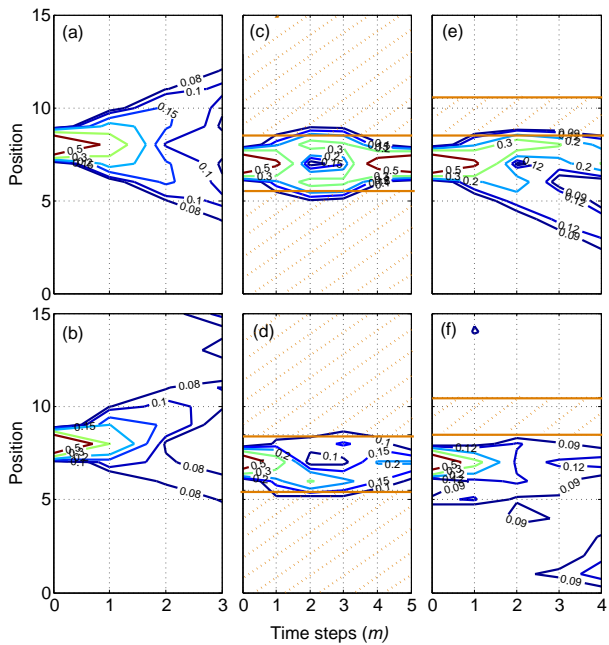}
\caption{
Theoretical (top row) and experimental (bottom row) contour plots showing
time evolution of spatial probabilities 
in the case of (a,b) free-particle, (c,d) particle in a well,
and (e,f) reflection from a barrier.
Here position is in units of $\Delta x$ and time is in units of $m\delta t$. 
In plots (c)-(f), the patched areas correspond to
non-zero values of potentials.
} \label{contour}
\end{figure}

\begin{figure}
\centering
\includegraphics[trim=0cm 6.3cm 0cm 6cm, clip=true,width=5.5cm]{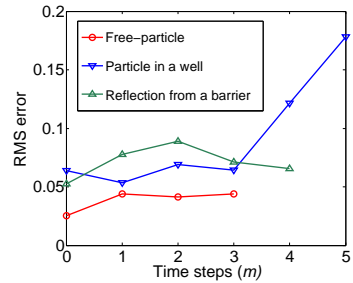}
\caption{
RMS error as a function of time-steps for the three potentials
discussed in the text.
}
\label{errors}
\end{figure}

In order to compare the overall experimental performance 
with theory, we use the contour plots of the spatial probability with evolution
time shown in Fig. \ref{contour} (a) and (b).  
Here the experimental contours (Fig. \ref{contour} (b)) are obtained by normalizing 
the sum of spectral intensities at each time-steps to unity, so that the effect of overall-decay is
removed.
Though the experimental contours
deviate from theory (Fig. \ref{contour} (a)), the 
over-all pattern appears to match, indicating a satisfactory quantum simulation of 
the free-particle case.  We use root-mean-square (RMS) errors to estimate the 
quality of the simulation (in a scale of 0 to 1).  
The RMS errors for various steps of evolution are 
plotted in Fig. \ref{errors}.  In the case of free-particle simulation, the maximum
RMS error was less than 0.05.

For the potential-well, the GRAPE evolution pulse had a lesser average fidelity of 0.95.
Fig. \ref{well} displays the results of simulations of particle in a well.  
Again the first trace is the reference and subsequent traces display the results
corresponding to various evolution intervals.  The initial state $\ket{j=7}$
is not a stationary state, but has a periodicity close to $5\delta t$.
Comparison with the expected probabilities (shaded bars), indicates a gradual
decay of the spectral lines, again due to decoherence. 
Comparison of the theoretical contours of probabilities in Fig. \ref{contour}(c) with experimental
contours in Fig. \ref{contour}(d) indicates a fairly good simulation of the dynamics.
Further, Fig. \ref{errors} indicates that RMS errors build up rapidly from the 4th step.
However the overall experimental contour patterns in Fig. \ref{contour}(d) 
reveal the $5\delta t$ periodicity as expected.

The GRAPE evolution pulse simulating reflection from a barrier 
also had a lesser average fidelity of 0.95.
Fig. \ref{bar} displays the corresponding results.  
Again the initial state $\ket{j=7}$ is not a stationary state.  As it spreads,
it gets reflected from the barrier and is constrained to one side of the barrier.
This phenomenon is evident in both the expected probabilities (shaded bars) as
well as in the experimental spectral lines.
The contours of probabilities in Fig. \ref{contour}(e) and (f) also show a 
similar behavior.  In this case, however, the pattern of experimental contours
does not match exactly with the theory.  This is again due to the lower fidelity of
the GRAPE pulse and other hardware limitations as mentioned earlier.  
The RMS errors (Fig. \ref{errors}) also indicate a poorer performance for the initial time-steps.

\section{Conclusions}
Simulation of quantum systems is an important motivation 
for building quantum processors.  It has been theoretically established
that such quantum processors, in which Hamiltonian can be controlled,
can simulate other quantum phenomena much more efficiently than
classical processors.  In this work we used a 5-qubit NMR quantum
simulator for simulating the dynamics of a quantum particle in
three one-dimensional potentials.  
The simulator consisted of a mutually interacting 5-spin system partially
oriented in a liquid crystal, and had stronger spin-spin
interactions and convenient spectral dispersions.
We used one of the qubits as the 
ancilla, which assisted in not only preparing the initial state,
but also direct spectral read-out of spatial probabilities at
various intervals of evolution.  The Schrodinger dynamics under various
potentials were implemented by robust RF pulses designed by GRAPE technique.
While the experimental results indicated an overall agreement with theory,
the deviations are mainly due to the temporal fluctuations in Hamiltonian
parameters of the 5-qubit register caused by local changes in the
order parameter of the liquid crystal.  Other sources of errors
included decoherence and spectrometer limitations.  
This work is an initial step in such direct quantum simulations
and needs several improvements towards achieving a versatile simulator.
The future work could include initializing wave-functions with 
any desired complex amplitude distributions, achieving higher 
spatial as well as temporal resolutions, and simulating with better fidelities.
With the increase in the size of quantum register and development of
control techniques, it may also be possible to simulate dynamics
of multi-dimensional potentials with higher complexities.

\section*{Acknowledgements}
The authors are grateful to Mr. Hemant Katiyar, Mr. Koteswara Rao, Mr. Abhishek Shukla,
Prof. Apoorva Patel, and Prof. Anil Kumar for discussions.  This work was partly supported by 
DST project SR/S2/LOP-0017/2009.

\bibliographystyle{apsrev4-1}
\bibliography{qsimuNotes12}

\end{document}